\documentclass[twocolumn]{article}
\usepackage{graphicx} 
\usepackage{amsmath}
\usepackage[hidelinks]{hyperref}

\title{Viral Evolution Under Physical Constraints: Decay, Mutation, and Transmission as a Constrained Optimization Problem}
\author{
Mohammad Rasoolinejad\\
Postdoctoral Researcher, Northwestern University\\
\texttt{Rasoolinejad@u.northwestern.edu}
}

\date{December 2025}

\begin{document}

\maketitle

\begin{abstract}
Viruses display striking diversity in structure, transmission mode, immune interaction, and evolutionary behavior. Despite this diversity, viral strategies are not unconstrained. Here we present a unifying framework that treats viral evolution as a problem of constrained optimization governed by physical decay, immune pressure, mutation robustness, and transmission architecture. We model virions as multi-component physical systems subject to irreversible environmental failure and viruses as replicators operating under immune-driven selection and mutation–selection balance. Within this framework, major viral transmission strategies arise as necessary solutions rather than taxonomic accidents. Environmentally transmitted and airborne viruses are predicted to be structurally simple, chemically stable, and reliant on replication volume rather than immune suppression. Structurally complex viruses tolerate rapid environmental decay by encoding immune-modulatory machinery, latency, or persistent replication, at the cost of reduced mutation robustness. Temperature-dependent seasonality emerges naturally from the thermally activated nature of viral decay, without invoking host behavior.
\end{abstract}

\section{Introduction}

Viruses are obligate intracellular parasites that depend entirely on host cellular machinery for replication while lacking autonomous metabolic processes, growth, or homeostasis. Outside the host, a virus exists as a virion; a physically inert infectious particle composed of a nucleic acid genome enclosed within a protein capsid and, in many cases, surrounded by a lipid envelope derived from host cell membranes \cite{Flint2015, KnipeHowley2021}. This minimal yet highly optimized structure places viruses at the boundary between biological and physical systems.

Traditional virus classification has emphasized molecular and morphological characteristics, including genome type (DNA or RNA), strandedness, capsid symmetry, presence or absence of an envelope, and replication strategy \cite{Gelderblom1996, Koonin2020, CaetanoAnolles2023}. While these criteria are indispensable for taxonomy, they provide limited insight into how viral structure constrains environmental persistence, transmission modality, immune interaction, and evolutionary strategy.

Environmental persistence has long been recognized as a central determinant of viral ecology and transmission \cite{Abad1994, Vasickova2010}. Non-enveloped (``naked'') viruses generally display greater resistance to physicochemical stressors such as desiccation, temperature variation, ultraviolet radiation, and extremes of pH. This robustness enables fecal--oral, fomite-mediated, and waterborne transmission routes \cite{Abad1994, Vasickova2010, Boone2020}. In contrast, enveloped viruses possess a lipid bilayer that is highly susceptible to environmental disruption, resulting in reduced survival outside the host and a stronger reliance on close contact, droplets, or aerosols for transmission \cite{Casanova2010, Firquet2015}.

A large experimental literature has quantified viral decay in air, on surfaces, and in aqueous environments. Across diverse virus families, loss of infectivity is often well approximated by first-order kinetics, with decay constants strongly dependent on temperature, humidity, surface type, and exposure to radiation \cite{Casanova2010, Firquet2015, Silverman2021}. Comparative studies consistently show that small, non-enveloped viruses exhibit substantially longer environmental half-lives than larger, enveloped respiratory viruses such as influenza viruses and coronaviruses \cite{Firquet2015, vanDoremalen2020, Zeng2023}.

At the same time, viral success within hosts is governed by an entirely different set of constraints. Once infection is established, viruses must replicate under intense selective pressure from innate and adaptive immune responses. Many viruses encode accessory proteins that interfere with immune signaling, antigen presentation, or apoptosis, thereby enhancing replication success at the cost of increased structural and genetic complexity \cite{FinlayMcFadden2006, KnipeHowley2021}. Others rely primarily on rapid replication and high population turnover to outrun immune containment, often generating extensive genetic diversity in the process \cite{Domingo2016, Lauring2010}.

High mutation rates, particularly in RNA viruses, give rise to quasispecies populations: dynamic clouds of related variants that collectively determine viral fitness \cite{Eigen1971, Domingo2012}. While genetic diversity enables rapid adaptation and immune escape, it also imposes an error threshold beyond which essential information is lost, constraining viable mutation rates \cite{Lauring2010, Ojosnegros2011}. Structural and functional complexity further tightens these constraints by reducing tolerance for mutational disruption.

Despite decades of research in molecular virology, environmental virology, and viral evolution, these domains remain largely disconnected. There is no unified theoretical framework that explicitly links virion structure, environmental decay, immune evasion strategy, mutation robustness, and transmission mode under a common set of physical and evolutionary constraints.

This paper seeks to bridge that gap. We propose a physically grounded framework in which viruses are treated as multi-component systems subject to irreversible environmental failure and immune-driven selection. By explicitly modeling decay, replication, mutation, and immune interaction, we show how viral diversity emerges as a constrained optimization problem governed by trade-offs between structural complexity, environmental persistence, and evolutionary adaptability.

\section{Structural Complexity, Failure Pathways, and Environmental Decay}

\subsection{Virions as Multi-Component Physical Systems}

A virion is best understood as a composite physical system composed of multiple essential components, each of which must remain intact for infectivity to be preserved. These components include the protein capsid, the viral genome, and, in enveloped viruses, a lipid membrane and associated surface glycoproteins. Additional accessory or matrix proteins may also be required to maintain structural integrity or mediate host entry \cite{Gelderblom1996, Flint2015}.

We represent a virion as a set of indispensable elements
\begin{equation}
\mathcal{V} = \{C, E, P_1,\dots,P_n, G\},
\end{equation}
where $C$ denotes the capsid, $E$ the lipid envelope when present, $P_i$ surface or structural proteins, and $G$ the genomic nucleic acid. Loss of infectivity occurs when any one of these elements undergoes irreversible degradation or functional failure.

Environmental stressors (including temperature fluctuations, desiccation, ultraviolet radiation, and chemical exposure) act independently or synergistically on these components. Non-enveloped viruses, characterized by rigid and highly symmetric capsids, generally exhibit greater resistance to such stresses, whereas enveloped viruses are vulnerable to lipid membrane disruption and glycoprotein denaturation \cite{Abad1994, Vasickova2010, Firquet2015}. This structural distinction provides a physical basis for differences in environmental persistence across virus families.

\subsection{Environmental Decay as a Reliability Process}

Loss of viral infectivity outside the host can be modeled using concepts from reliability theory, in which a system fails when any one of its essential components fails. Each virion component $i \in \mathcal{V}$ is associated with a hazard rate $h_i$, defined as the instantaneous probability per unit time that the component undergoes irreversible damage under specified environmental conditions \cite{Nelson2004}.

Assuming that component failures are approximately independent and rare on short timescales, the effective system-level hazard rate $\lambda$ governing loss of infectivity is given by the sum of individual hazard rates:
\begin{equation}
\lambda = \sum_{i \in \mathcal{V}} h_i.
\end{equation}

This additive structure reflects a competing-risks process: infectivity is lost when the first essential component fails. Under these assumptions, the survival probability of an infectious virion follows an exponential law, consistent with first-order decay kinetics widely observed in experimental studies of virus survival in air, on surfaces, and in aqueous media \cite{Casanova2010, Firquet2015, Silverman2021}.

Let $N(t)$ denote the number of infectious virions remaining at time $t$ after release into the environment. The decay dynamics are then described by
\begin{equation}
\frac{dN}{dt} = -\lambda N,
\end{equation}
with solution
\begin{equation}
N(t) = N_0 e^{-\lambda t},
\end{equation}
where $N_0$ is the initial number of infectious particles.

\subsection{Half-Life and Environmental Persistence}

The environmental half-life $t_{1/2}$ is defined as the time required for $N(t)$ to decrease to $N_0/2$. From the exponential decay law,
\begin{equation}
t_{1/2} = \frac{\ln 2}{\lambda}.
\end{equation}

This expression provides a direct quantitative link between structural fragility and persistence. Viruses with fewer vulnerable components or more robust molecular architectures exhibit smaller $\lambda$ and therefore longer half-lives. Empirical measurements confirm that non-enveloped viruses can remain infectious on surfaces for weeks, whereas many enveloped viruses lose viability within hours to days under comparable conditions \cite{Abad1994, Firquet2015, vanDoremalen2020}.

Importantly, half-life is not an intrinsic constant of a virus alone but a function of environmental context. The same virion may exhibit dramatically different decay constants depending on temperature, humidity, surface composition, and exposure to radiation.

\subsection{Environmental Drivers of Component Hazard Rates}

Environmental variables modulate the hazard rates $h_i$ by altering the stability of molecular structures. Elevated temperatures accelerate protein unfolding, lipid phase transitions, and nucleic acid damage, thereby increasing $\lambda$ \cite{Casanova2010, Anderson2023}. Ultraviolet radiation induces direct genomic damage and protein cross-linking, contributing additional failure pathways \cite{Lytle2005}.

Relative humidity exerts more complex effects. Low humidity can preserve infectivity for some airborne viruses by limiting hydrolytic damage, whereas intermediate humidity may accelerate decay through enhanced solute concentration and membrane stress \cite{Yang2012, Morris2021}. Surface properties further influence decay by affecting desiccation dynamics and the formation of protective microenvironments \cite{Zeng2023}.

Taken together, these factors determine the aggregate decay constant $\lambda$ as an emergent property of both virion structure and environmental context. Structural complexity increases the number of potential failure modes, while environmental conditions regulate the rates at which those modes are activated. This framework explains why environmental persistence varies systematically across virus classes and provides a physical basis for incorporating decay into transmission and evolutionary models.

\section{Replication, Immune Interaction, and Mutation Robustness}

\subsection{Replication Success Under Immune Pressure}

Once a virus enters a host cell, environmental decay is no longer the dominant selective force. Instead, viral replication proceeds under intense pressure from innate and adaptive immune responses, including interferon signaling, restriction factors, neutralizing antibodies, and cytotoxic lymphocytes \cite{Murphy2017, Janeway2021}. Viral fitness within the host therefore depends on both intrinsic replication capacity and the ability to suppress, evade, or tolerate immune attack.

We represent viral replication success by an effective reproductive output $R$, determined by an intrinsic replication rate $r$ modulated by immune pressure. Let $I$ denote the net immune pressure exerted on the virus, and let $K$ represent the cumulative immune evasion or suppression capacity encoded by the virus. A phenomenological expression for replication success is
\begin{equation}
R = r \, e^{-\beta (I - K)},
\end{equation}
where $\beta$ sets the sensitivity of replication to immune pressure. When $K \ge I$, immune effects are partially neutralized and replication proceeds efficiently; when $I \gg K$, replication is rapidly curtailed.

This formulation captures a central biological reality: immune interaction is competitive rather than purely inhibitory. Viruses that encode proteins interfering with antigen presentation, interferon signaling, or apoptosis effectively reduce the net immune pressure they experience \cite{FinlayMcFadden2006, RandallGoodbourn2008}. Such capabilities, however, require additional genetic material and regulatory complexity.

\subsection{Mutation, Replication Fidelity, and Quasispecies}

Viral replication fidelity strongly influences evolutionary dynamics. Many RNA viruses replicate using virus-encoded RNA-dependent RNA polymerases that lack proofreading activity, resulting in mutation rates orders of magnitude higher than those of cellular DNA replication \cite{Drake1999, Peck2018}. Consequently, viral populations do not exist as single genotypes but as distributions of closely related variants known as quasispecies \cite{Eigen1971, Domingo2012}.

Within the quasispecies framework, selection acts on the population as a whole rather than on isolated genotypes. Genetic diversity enables rapid exploration of sequence space, facilitating immune escape, host adaptation, and resistance to antiviral pressures \cite{Lauring2010, Domingo2016}. However, excessive mutation introduces a risk of error catastrophe, in which essential genetic information is lost faster than it can be regenerated by selection \cite{Eigen1971, Ojosnegros2011}.

Let $\mu$ denote the per-replication mutation rate. Viral fitness depends non-monotonically on $\mu$: increasing mutation initially enhances adaptability, but beyond a critical threshold fitness declines sharply due to accumulation of deleterious mutations.

\subsection{Mutation Robustness and Structural Constraints}

Not all viruses tolerate mutation equally. We define mutation robustness $\rho$ as the probability that a random mutation preserves functional infectivity. Mutation robustness depends on genome organization, protein folding constraints, and the degree of structural and functional coupling between viral components.

Viruses with compact, modular architectures often tolerate higher mutation rates because many mutations are neutral or nearly neutral. In contrast, structurally complex viruses encode proteins with tightly constrained interactions, reducing $\rho$ and limiting viable mutational space \cite{Wylie2014, Sanjuan2010}.

We model mutation robustness as a decreasing function of structural complexity $C_s$:
\begin{equation}
\rho = f(C_s), \qquad \frac{d\rho}{dC_s} < 0.
\end{equation}

This relationship reflects the intuitive but critical constraint that complexity increases fragility. As more components and interactions are required for functionality, fewer mutations can be tolerated without catastrophic loss of infectivity.

\subsection{Effective Fitness Under Replication--Mutation Trade-Offs}

To unify replication, immune pressure, and mutation, we define a composite within-host fitness $\mathcal{F}$ as
\begin{equation}
\mathcal{F} = R \times g(\mu,\rho),
\end{equation}
where $g(\mu,\rho)$ captures the net evolutionary benefit of mutation under robustness constraints.

A simple functional form consistent with quasispecies theory is
\begin{equation}
g(\mu,\rho) = \rho \mu e^{-\gamma \mu},
\end{equation}
where $\gamma$ controls the severity of error accumulation. This expression increases with mutation rate at low $\mu$, reflecting adaptive diversification, but declines at high $\mu$ as deleterious mutations dominate.

This formulation reproduces several well-established empirical patterns. RNA viruses with high mutation robustness operate near the peak of $g(\mu,\rho)$, maintaining mutation rates close to the error threshold. DNA viruses with complex architectures operate at much lower $\mu$, relying instead on immune suppression and regulatory control to sustain replication \cite{Sanjuan2010, Domingo2016}.

\subsection{Interpretation and Evolutionary Strategy}

Within this framework, viral strategies emerge naturally from physical and biological constraints:

\begin{itemize}
  \item Viruses with high mutation robustness and limited immune suppression rely on rapid replication and genetic diversity to maintain fitness under immune pressure \cite{Lauring2010}.
  \item Viruses with low mutation robustness compensate by encoding immune-modulatory proteins, trading adaptability for controlled replication and antigenic stability \cite{FinlayMcFadden2006}.
\end{itemize}

Importantly, mutation strategy cannot be chosen independently of structure. Structural complexity simultaneously increases immune evasion capacity and decreases tolerance for mutation. Viral evolution therefore reflects a constrained optimization problem rather than unconstrained adaptation.

This section establishes the within-host component of viral fitness. In the following section, we integrate this replication–mutation framework with environmental decay to define total transmission fitness and derive a physically grounded viral classification.

\section{Unified Transmission Fitness and Viral Classification}

Viral success requires survival across two fundamentally distinct and sequential domains. In the external environment, virions exist as inert physical objects subject to irreversible degradation driven by physicochemical stressors. Within the host, viruses become active replicators engaged in competition with host immune defenses and constrained by mutation--selection balance. These two stages impose opposing selective pressures. Environmental survival favors structural simplicity, chemical robustness, and minimal failure pathways. In contrast, within-host replication favors molecular complexity, immune modulation, and regulatory control. A complete measure of viral success must therefore integrate both environmental persistence and replication success.

\subsection{Definition of Total Transmission Fitness}

We define the total transmission fitness $\Phi$ of a virus as the expected number of successful secondary infections generated per released virion. This quantity depends on (i) the probability that a virion remains infectious long enough to encounter a host and (ii) the probability that replication succeeds once infection is established.

From Section~2, environmental survival follows first-order decay governed by the system-level hazard rate $\lambda$:
\begin{equation}
P_{\text{env}}(t) = e^{-\lambda t}.
\end{equation}

From Section~3, within-host replication success is captured by the effective reproductive output $R$, incorporating intrinsic replication rate, immune pressure, immune evasion, and mutation constraints.

The total transmission fitness is then
\begin{equation}
\Phi = \int_0^\infty P_{\text{env}}(t)\, R \, dt.
\end{equation}

Assuming that $R$ is independent of environmental residence time once infection occurs, the integral evaluates to
\begin{equation}
\Phi = \frac{R}{\lambda}.
\end{equation}

This expression makes explicit the central trade-off governing viral evolution: transmission fitness increases with replication success and immune evasion but decreases with environmental fragility.

\subsection{Structural Complexity as a Control Parameter}

Structural complexity enters the fitness function primarily through the decay constant $\lambda$. As shown in Section~2, $\lambda$ is the sum of hazard rates associated with essential virion components:
\begin{equation}
\lambda = \sum_{i \in \mathcal{V}} h_i.
\end{equation}

Increasing structural complexity introduces additional components, interfaces, and fragile interactions, thereby increasing the number of failure pathways and raising $\lambda$. Conversely, complexity enhances immune evasion capacity and regulatory control, increasing $R$.

Viral evolution therefore operates under a constrained optimization problem:
\begin{equation}
\max \; \Phi = \max \left( \frac{R(C_s)}{\lambda(C_s)} \right),
\end{equation}
where $C_s$ denotes structural complexity. Neither extreme simplicity nor maximal complexity is universally optimal; instead, viable viral strategies occupy constrained regions of this trade-off space.

\subsection{Emergent Viral Classes}

Distinct viral transmission strategies emerge naturally from different optima of $R/\lambda$:

\begin{itemize}
  \item \textbf{Simple, environmentally stable viruses} achieve low $\lambda$ through minimal structure and chemical robustness. Their replication strategy relies on rapid population expansion and genetic diversity rather than immune suppression.
  \item \textbf{Complex, close-contact viruses} tolerate higher $\lambda$ by compensating with increased $R$, achieved through immune modulation, latency, or persistent replication.
\end{itemize}

These classes are not arbitrary taxonomic groupings but emergent solutions to a physical optimization problem imposed by environmental decay and immune selection.

\subsection{Implications for Viral Evolution}

The expression $\Phi = R/\lambda$ implies that viral evolution is constrained by a conservation-like principle: gains in immune evasion and replication capacity must be paid for with increased environmental fragility, and vice versa. No virus can simultaneously maximize $R$ and minimize $\lambda$ without bound.

This constraint explains several empirical observations:
\begin{itemize}
  \item Airborne viruses tend toward structural simplicity and environmental stability.
  \item Highly complex viruses rely on close contact, latency, or persistent infection.
  \item Vaccines are durable primarily for viruses constrained to low mutation rates by structural complexity.
\end{itemize}

The framework therefore provides a unifying lens for understanding viral diversity, transmission modes, and long-term controllability.

\section{Effect of Temperature on Survival and Spread Dynamics}

Environmental temperature plays a central role in governing the survival of physical and biological agents outside their active operating domains. For viruses, temperature primarily affects transmission by modulating the rate of irreversible decay processes acting on virion components, including protein denaturation, lipid membrane disruption, and nucleic acid damage. Because these processes are thermally activated, their rates exhibit strong temperature dependence that can be captured using well-established physical chemistry principles.

\subsection{Temperature Dependence of Viral Decay}

Empirical studies across virus families consistently show that viral inactivation rates increase with temperature and decrease under colder conditions \cite{Casanova2010, Firquet2015, Morris2021}. This behavior is well described by the Arrhenius equation, which relates the rate constant of a thermally activated process to absolute temperature:
\begin{equation}
k(T) = A \exp\left(-\frac{E_a}{RT}\right),
\end{equation}
where $k(T)$ is the rate constant at temperature $T$, $A$ is the pre-exponential factor, $E_a$ is the activation energy, and $R$ is the universal gas constant \cite{Atkins2014}.

For viral environmental decay, the rate constant $k(T)$ corresponds to the system-level decay constant $\lambda(T)$ introduced in Section~2. Substituting, environmental survival follows
\begin{equation}
N(t;T) = N_0 \exp\!\left[-\lambda(T)t\right],
\end{equation}
with
\begin{equation}
\lambda(T) = A \exp\!\left(-\frac{E_a}{RT}\right).
\end{equation}

\subsection{Temperature Dependence of Environmental Half-Life}

The environmental half-life $t_{1/2}(T)$ is given by
\begin{equation}
t_{1/2}(T) = \frac{\ln 2}{\lambda(T)}.
\end{equation}
Substituting the Arrhenius form yields
\begin{equation}
t_{1/2}(T) = \ln 2 \cdot A^{-1} \exp\!\left(\frac{E_a}{RT}\right).
\end{equation}

This expression demonstrates that decreasing temperature produces an exponential increase in viral half-life. Even modest reductions in ambient temperature can therefore yield large increases in environmental persistence, particularly for structurally simple viruses with low baseline decay constants.

This physical mechanism provides a direct explanation for the enhanced survival of many viruses in cold environments and under refrigerated or winter conditions, independent of host behavior or immune dynamics \cite{Lowen2007, Marr2019}.

\subsection{Survival Time and Effective Contact Probability}

For viruses whose transmission depends on environmental persistence—such as airborne or fomite-mediated viruses—the probability of successful host contact increases with survival time. Let $S(t;T)=\exp[-\lambda(T)t]$ denote the survival probability of a virion at time $t$. If effective contacts occur at a rate proportional to survival, the cumulative probability of contact is
\begin{equation}
P_{\text{contact}}(T) \propto \int_0^\infty S(t;T)\, dt = \frac{1}{\lambda(T)}.
\end{equation}

Thus, colder temperatures increase transmission probability by reducing $\lambda(T)$, extending survival time, and enlarging the window during which infectious contacts can occur. This effect is purely physical and does not rely on changes in host susceptibility or behavior.

\subsection{Temperature-Driven Modulation of Growth Rate}

In epidemiological terms, the effective reproduction rate $\beta(T)$ scales inversely with the decay constant:
\begin{equation}
\beta(T) \propto \frac{1}{\lambda(T)} = A^{-1} \exp\!\left(\frac{E_a}{RT}\right).
\end{equation}

The net growth rate $r(T)$ of infections can be expressed as
\begin{equation}
r(T) = \beta(T) - \delta,
\end{equation}
where $\delta$ represents removal processes such as immune clearance or recovery.

Because $\beta(T)$ increases exponentially as temperature decreases, the system may transition from decay ($r<0$) to growth ($r>0$) solely due to temperature change. This provides a mechanistic explanation for seasonal outbreaks driven by environmental persistence rather than intrinsic changes in viral replication or virulence.

\subsection{Scope and Applicability}

The temperature–decay formulation developed here is most directly applicable to viruses whose transmission is constrained by environmental survival, particularly airborne and environmentally mediated viruses. In these systems, decay outside the host constitutes the dominant bottleneck, allowing transmission dynamics to be approximated by first-order survival kinetics.

For structurally complex viruses transmitted primarily through close or direct contact, temperature influences transmission through multiple interacting pathways, including host behavior, immune response, and intracellular replication efficiency \cite{Foxman2015}. In such cases, environmental decay represents only one component of a multifactorial process, and the present formulation should be interpreted as isolating the physical persistence contribution rather than providing a complete transmission model.

Nonetheless, for airborne viruses characterized by structural simplicity and environmental robustness, temperature-driven modulation of decay provides a sufficient and predictive explanation for observed seasonality and cold-weather transmission enhancement.

\section{Transmission-Based Classification and Evolutionary Trade-Offs}

The diversity of viral life cycles can be understood as a constrained response to a small number of fundamental physical and biological trade-offs. Rather than representing arbitrary evolutionary outcomes, observed viral strategies emerge as solutions to a common optimization problem: maximizing total transmission fitness under simultaneous constraints imposed by environmental decay, immune pressure, mutation robustness, and host viability.

The unified fitness expression derived in Section~4,
\begin{equation}
\Phi = \frac{R}{\lambda},
\end{equation}
makes explicit that viral success depends not on replication or persistence alone, but on their ratio. Any increase in replication success $R$ achieved through added molecular complexity necessarily increases structural fragility and thus the environmental decay constant $\lambda$. Conversely, minimizing $\lambda$ by simplifying structure restricts immune evasion and regulatory control, limiting achievable $R$. Viral evolution therefore proceeds along a narrow feasible manifold rather than an unconstrained adaptive landscape.

A natural first division arises between viruses whose transmission is bottlenecked primarily by environmental survival and those whose transmission is bottlenecked by host-to-host contact. Environmentally mediated viruses, including airborne viruses, must remain infectious outside the host long enough to encounter new hosts. For these viruses, the dominant selective pressure acts prior to host entry, favoring structural simplicity, chemical robustness, and resistance to environmental stressors. Lipid envelopes, large accessory proteins, and complex regulatory machinery are strongly disfavored because they introduce additional failure pathways and raise $\lambda$. As a result, environmentally transmitted viruses tend to be non-enveloped or minimally structured and occupy regions of parameter space characterized by low structural entropy and low environmental decay.

The consequence of this constraint is that environmentally transmitted viruses possess limited capacity for immune suppression once inside the host. Instead, their primary strategy is temporal: upon infection, they exploit the short delay before effective immune responses are established to enter a rapid exponential replication phase. High replication volume compensates for limited immune modulation by increasing transmission probability and generating population-level genetic diversity. Mutation effectiveness in this regime is achieved not through per-replication error rates alone, but through scale. Large viral populations combined with repeated host-to-host passage allow selection to operate efficiently across many replication cycles, even when individual transcription events are relatively accurate. Cold environments amplify this strategy by reducing environmental decay, effectively lowering $\lambda$ and expanding the transmission window. Seasonality and cold-weather preference therefore emerge naturally from physical persistence rather than from enhanced biological fitness within hosts.

Viruses that rely primarily on close or direct contact occupy a fundamentally different region of the trade-off space. Because environmental persistence is no longer the dominant bottleneck, these viruses can tolerate higher decay constants and thus greater structural complexity. This relaxation permits the encoding of immune-modulatory proteins, regulatory networks, and replication control mechanisms that enhance within-host fitness. However, increased complexity constrains mutation robustness, reducing the range of tolerable genomic variation and slowing antigenic evolution.

Within the close-contact regime, multiple evolutionary strategies emerge depending on how viruses resolve the tension between replication speed, immune pressure, and host survival. Some viruses prioritize replication volume in a manner analogous to environmentally transmitted viruses, entering immediate exponential growth before immune containment. Because transmission does not depend on environmental survival, this strategy can succeed even with fragile virions. The cost, however, could be severe collateral damage to host tissues, if involving vital organs, leading to acute disease and elevated host mortality. Transmission success in this regime is front-loaded in time and incompatible with long-term host survival.

Other viruses adopt a delayed replication strategy, minimizing early transcriptional activity to reduce immune detection before transitioning into exponential growth. This approach extends the window for establishment but still incurs substantial pathology once replication accelerates. A more conservative solution is achieved through latency or dormancy, in which viral activity is restricted to avoid immune clearance, punctuated by episodic reactivation events. By decoupling persistence from continuous replication, these viruses maximize host longevity and generate multiple transmission opportunities over extended timescales.

At the extreme end of structural and regulatory sophistication are viruses that maintain ongoing replication while persisting within the host, achieved through genome integration or continuous immune modulation. In this regime, immune pressure becomes the dominant selective force, driving sustained within-host evolution. Transmission is no longer a single event but an emergent property of long-term coexistence with the host. The cost of this strategy is extreme dependence on complex molecular machinery and tight constraints on mutation robustness.

Across all regimes, mutation strategy remains subordinate to structural constraints. High mutation rates are viable only when robustness permits; immune escape through genetic variation is feasible only within narrow regions of the genome. No virus can simultaneously achieve high mutation rates, extensive immune evasion, and environmental stability. Each observed viral strategy reflects a compromise enforced by physical decay, biochemical fragility, and evolutionary timing.

This perspective reframes viral classification not as a catalog of families or genomes, but as a map of constrained solutions to a shared optimization problem. Transmission modality, structural complexity, immune interaction, and mutation capacity are not independent traits but interlocked variables shaped by fundamental trade-offs. Understanding these constraints clarifies why certain viruses are controllable by vaccination, why others exhibit persistent antigenic drift, and why environmental interventions are effective only for specific classes of pathogens.

\section{Mutation, Selection, and Transmission Under Transcriptional Constraints}

Mutation alone does not guarantee viral evolution. For a mutation to contribute to long-term viral fitness, three conditions must be simultaneously satisfied: mutations must be generated at a sufficient rate, natural selection must act efficiently on the resulting variants, and selected variants must be successfully transmitted to subsequent hosts. Failure of any one of these conditions renders mutation evolutionarily ineffective, regardless of its molecular origin.

A fundamental constraint on viral mutation arises from the fidelity of transcription and replication machinery. Host DNA and RNA polymerases operate with exceptionally low error rates and often incorporate proofreading and repair mechanisms. Viruses that rely predominantly on host transcription therefore inherit these low mutation rates, limiting the generation of genetic diversity per replication cycle \cite{Drake1999, Sanjuan2010}. In such systems, meaningful variation can arise only through very large numbers of replication events.

This constraint immediately distinguishes viral strategies. Environmentally transmitted viruses, particularly airborne viruses, overcome transcriptional fidelity not by increasing per-cycle mutation rates but by exploiting scale. High replication volume within hosts, combined with the release of large numbers of virions into the environment, generates vast mutational ensembles even when individual replication events are accurate. Selection then occurs externally: environmental stressors and immune defenses of recipient hosts act as powerful filters that eliminate nonviable variants before or at the point of infection. In this regime, mutation effectiveness is achieved through high-throughput replication and high-throughput selection. Transmission bottlenecks are wide, allowing large numbers of variants to pass between hosts. As a result, advantageous mutations arising anywhere in the replication process have a reasonable probability of contributing to future viral populations. Mutation is therefore evolutionarily productive despite strict transcriptional fidelity.

Close-contact viruses face a fundamentally different evolutionary landscape. Transmission events typically involve small inocula, producing narrow bottlenecks that drastically reduce the number of variants delivered to the next host \cite{McCrone2018}. Under these conditions, many mutations—regardless of their potential advantage—are lost simply because they are not transmitted. Natural selection occurring late in infection is therefore ineffective at shaping long-term evolution unless variants arise early enough to dominate the population before transmission.

This creates a timing problem, for mutation to contribute meaningfully to long-term viral fitness, selection must act before or during the earliest stages of infection. Viruses transmitted through close contact cannot rely on extremely large replication volumes to generate sufficient genetic diversity prior to transmission. Because transmission bottlenecks are narrow, most late-arising variants are lost regardless of their potential advantage. To overcome this limitation, such viruses must increase the rate at which diversity is generated early in infection and trust host's immune system to do the selection. One solution is the use of virus-encoded polymerases with reduced replication fidelity. By decoupling mutation generation from the high accuracy of host transcription machinery, viruses can elevate mutation rates per replication cycle and produce genetically diverse populations even at relatively low replication volumes \cite{Peck2018}. 

This strategy, however, is viable only when mutation robustness is sufficiently high to tolerate increased error rates without compromising essential structure or function. Structural complexity imposes strict limits on this approach. Viruses encoding tightly coupled protein networks, elaborate capsids, or multiple regulatory proteins exhibit low mutation robustness; even small genetic perturbations can disrupt essential functions. For these viruses, increasing mutation rate risks immediate loss of viability. Consequently, many structurally complex viruses maintain low mutation rates and rely on immune modulation, latency, or persistence rather than rapid genetic adaptation.

Dormancy and latency offer partial but limited solutions to the selection timing problem. By minimizing transcriptional activity, viruses can evade immune clearance and extend host survival. However, latency often suppresses replication and mutation simultaneously, slowing evolutionary change rather than accelerating it. Only in specific cases where limited replication occurs during persistence, does latency contribute meaningfully to diversification. Thus, dormancy should be viewed primarily as a strategy for immune avoidance and transmission longevity, not as a general mechanism for enhancing mutation effectiveness.

Persistent replicators represent an extreme solution. By maintaining ongoing replication under continuous immune pressure, these viruses convert the host immune system into a within-host selective environment. Selection acts continuously rather than episodically, allowing advantageous variants to rise even under narrow transmission bottlenecks. This strategy requires extensive immune modulation and precise regulatory control and is therefore accessible only to highly specialized viruses.

Across all regimes, mutation strategy is subordinate to physical and biological constraints. High mutation rates are advantageous only when supported by sufficient robustness, early selection, and effective transmission. Conversely, low mutation rates do not preclude evolution when replication scale and selection efficiency are high. Viral diversity thus reflects not differences in evolutionary ambition but differences in what mutation strategies are physically and biologically feasible.

\section{Comparative Case Studies Across the Transmission Landscape}

The hazard-based framework developed in this work predicts that viral strategies cluster into constrained regions of parameter space defined by environmental decay ($\lambda$), effective replication success within hosts ($R$), mutation rate ($\mu$), mutation robustness ($\rho$), and structural complexity as reflected in the number and fragility of essential components. To illustrate how these constraints manifest in real systems, we examine six representative viruses spanning environmentally transmitted, close-contact, latent, and persistent replicative regimes: norovirus, poxviruses (including smallpox), influenza virus, herpesviruses, and human immunodeficiency virus (HIV). These case studies are not intended as exhaustive descriptions, but as demonstrations that diverse viral behaviors emerge naturally from the same underlying trade-offs between environmental failure, replication timing, mutation generation, and immune interaction.

\paragraph{Norovirus: Extreme Environmental Stability and Low Failure Rates}

Noroviruses exemplify the simple, environmentally stable regime. They are non-enveloped, possess small and highly symmetric capsids, and lack auxiliary enzymatic machinery beyond what is strictly required for replication. As a result, the number of environmental failure pathways is minimal, and the aggregate environmental decay constant $\lambda$ is correspondingly small. Empirical studies consistently show that noroviruses remain infectious for prolonged periods on surfaces and in water, resisting detergents, desiccation, and temperature variation \cite{Vasickova2010, Boone2020, Estes2019}. Within hosts, noroviruses exhibit limited immune suppression and instead rely on rapid replication and large population sizes to sustain transmission. High mutation rates generate genetic diversity at the population level, enabling immune escape across hosts rather than within individual infections \cite{Domingo2016}. In this regime, transmission fitness is maximized primarily by minimizing $\lambda$ rather than by increasing $R$ through immune modulation. Cold environments further amplify transmission by reducing environmental hazard rates and extending survival time.

\paragraph{Poxviruses: Structural Complexity and Controlled Failure}

Poxviruses occupy a contrasting region of the hazard landscape. These large DNA viruses possess elaborate capsids and encode extensive molecular machinery for immune modulation, replication control, and host interaction. Structural complexity increases the number of potential environmental failure modes, resulting in larger decay constants and reliance on close-contact transmission. Mutation rates are low, and antigenic profiles remain remarkably stable over time. Rather than relying on mutation-driven immune escape, poxviruses suppress immune responses through encoded proteins that interfere with host signaling pathways \cite{FinlayMcFadden2006}. This strategy reflects a deliberate trade-off: increased replication success $R$ within hosts is achieved at the cost of increased environmental fragility and reduced mutation robustness.

Variola virus (smallpox) represents an extreme but instructive case within the poxvirus family. Despite significant virulence and efficient transmission through close contact, the virus exhibited minimal antigenic evolution over centuries. The amount of viral material transferred between hosts was typically limited, producing narrow transmission bottlenecks. Following infection, viral replication entered a rapid exponential phase, during which population expansion outpaced immune-mediated or competitive selection. As a result, mutations arising during replication had a vanishingly small probability of being both selectively advantageous and present early enough to dominate the viral population prior to transmission.

Structural and regulatory complexity further constrained mutation robustness, forcing the per-replication mutation rate $\mu$ to remain low and sharply limiting viable genetic variation \cite{Fenner1988}. In the absence of large transmission volumes or an effective early selection mechanism, the probability that beneficial mutations would both arise and be successfully transmitted was extremely small. Once global vaccination imposed uniform immune pressure, the virus lacked viable evolutionary pathways to evade immunity without compromising essential functions. From the hazard-based perspective, smallpox occupied a narrow region of parameter space in which replication success $R$ was sufficient for transmission, but limited mutation rate $\mu$ and low robustness $\rho$ made adaptive escape effectively inaccessible, leading ultimately to eradication.

\paragraph{Influenza Virus: Operation Near the Failure--Mutation Boundary}

Influenza viruses occupy an intermediate, adaptive regime. They are enveloped and structurally more complex than noroviruses, resulting in higher environmental decay rates and strong seasonality. At the same time, influenza viruses maintain mutation rates high enough to permit antigenic drift in surface glycoproteins, particularly hemagglutinin and neuraminidase \cite{Grenfell2004, Webster1992}. This strategy places influenza near a critical boundary: mutation rates are sufficiently high to evade population immunity, yet constrained enough to avoid widespread loss of structural or functional integrity. Selection acts both within hosts and across transmission cycles through repeated seasonal outbreaks. The hazard-based framework explains why influenza vaccines must be continually updated and why eradication remains infeasible despite sustained immunological pressure.

\paragraph{Herpesviruses: Latency as Hazard Avoidance}

Herpesviruses adopt a fundamentally different solution to the trade-off problem. These large, enveloped DNA viruses establish long-term latency in specific host cell types, maintaining minimal transcriptional and replicative activity to avoid immune clearance. Structural complexity is high, environmental fragility is substantial, and mutation rates are low \cite{Whitley1998}. Rather than maximizing replication or mutation, herpesviruses minimize effective hazard exposure by suppressing replication for extended periods. Episodic reactivation produces localized replication and limited transmission events without overwhelming host defenses. In this regime, transmission fitness is achieved through longevity and repeated low-intensity transmission opportunities rather than through rapid replication or environmental persistence.

\paragraph{HIV: Persistent Replication Under Continuous Selection}

Human immunodeficiency virus represents the most extreme adaptive regime considered here. HIV integrates into the host genome and maintains ongoing replication under continuous immune pressure. Although structurally complex and environmentally fragile, HIV compensates by employing virus-encoded polymerases with low fidelity, generating extraordinarily high mutation rates \cite{Nowak1996, Perelson1996}. In this regime, selection acts continuously within hosts rather than primarily between hosts. The immune system itself becomes the dominant selective environment, allowing advantageous variants to emerge even under narrow transmission bottlenecks. Transmission fitness arises from long-term persistence and diversification rather than from environmental survival or short-term replication bursts.

\paragraph{Synthesis Across Cases}

Across all cases, viral strategies align with the constraints predicted by the hazard-based framework. Norovirus maximizes fitness by minimizing environmental failure; poxviruses and smallpox invest in immune modulation at the cost of adaptability; influenza balances mutation and structural fragility near a critical boundary; herpesviruses avoid hazard through latency; and HIV converts immune pressure into a continuous engine of within-host evolution. These patterns are not contingent historical accidents. They are the necessary outcomes of optimization under environmental decay, structural fragility, mutation robustness, and selection timing constraints. 

\section{Implications for Public Health, Vaccination, and Long-Term Control}

The hazard-based framework developed in this work reframes public health intervention as a problem of altering dominant failure and selection pathways rather than applying a universal set of tools. Because viral transmission fitness is governed by the ratio, interventions succeed only when they act directly on the limiting factor for a given virus. Measures that fail to modify the dominant hazard or selection bottleneck may reduce short-term transmission but cannot produce durable control.

Durable vaccination is achievable only when immune pressure acts before the viral population enters rapid exponential growth and when adaptive escape is constrained by low mutation robustness and narrow transmission bottlenecks. Viruses with high structural and regulatory complexity often satisfy these conditions. Limited transmitted viral load, early exponential replication, and tightly constrained mutation rates sharply reduce the probability that beneficial mutations will both arise early and be transmitted. In such systems, vaccination does not merely suppress transmission temporarily but removes viable evolutionary pathways altogether, as demonstrated historically for smallpox.

By contrast, viruses with high mutation rates or early diversity generation cannot be eliminated through vaccination alone. When mutation is prevalent, either through low-fidelity polymerases or persistent replication, immune pressure selects escape variants rather than terminating transmission. In these cases, vaccination remains valuable for reducing disease severity, hospitalization, and mortality, but cannot be expected to eliminate circulation. The need for repeated vaccine reformulation or boosting reflects structural constraints on mutation and selection timing rather than shortcomings in immunological design.

An important implication of the hazard-based framework is the existence of hard evolutionary ceilings. Viruses cannot indefinitely reduce environmental decay without structural simplification, nor can they indefinitely increase mutation rates without triggering catastrophic loss of function. These limits explain why immune escape repeatedly converges on a narrow set of surface-exposed regions, while core structural and enzymatic components remain conserved across decades of evolution.

From a surveillance perspective, viral evolution should be interpreted as movement within a constrained hazard–selection space. Rapid increases in mutation rate without corresponding changes in transmission architecture may signal proximity to mutational collapse. Conversely, long-term antigenic stability combined with narrow transmission bottlenecks indicates susceptibility to durable vaccination. Monitoring mutation timing, transmission volume, and environmental stability therefore provides insight that sequence data alone cannot capture.

The framework also clarifies why many interventions appear effective only transiently. Measures that reduce transmission without altering the dominant hazard or selection bottleneck merely displace the system temporarily. Once the intervention is relaxed, the virus returns to its equilibrium region of parameter space. Durable control requires forcing the virus into a region where no viable combination of replication success, mutation generation, and hazard avoidance exists. The framework cautions against overgeneralization. No single intervention strategy is universally optimal. Environmentally stable viruses demand environmental hazard amplification. Structurally complex viruses with constrained mutation are susceptible to vaccination. Highly adaptive viruses require layered mitigation, continuous surveillance, and acceptance of long-term coexistence rather than eradication. These outcomes are not policy choices but consequences of physical limits, biological fragility, and evolutionary timing.

\section{Limitations and Extensions of the Framework}

The hazard-based framework developed in this work is intentionally coarse-grained. Its goal is not to reproduce detailed molecular mechanisms or to generate precise epidemiological forecasts, but to identify the dominant physical and biological constraints that bound viral strategies across transmission modes, environments, and evolutionary timescales. As a result, several limitations arise from necessary simplifications rather than from conceptual shortcomings.

A primary limitation is the aggregated treatment of virion structure. Structural complexity is represented implicitly through the number and fragility of essential components contributing to the aggregate environmental decay constant $\lambda$. In reality, virion failure arises from specific molecular events, including protein denaturation, lipid membrane disruption, capsid instability, and nucleic acid damage. The framework does not resolve these mechanisms individually, but instead assumes that their combined effect can be captured by an effective hazard rate. This abstraction enables generality across virus families but necessarily obscures virus-specific molecular pathways that may dominate under particular environmental conditions.

A second limitation concerns mutation dynamics. Mutation is characterized using two effective parameters: the mutation rate per replication cycle $\mu$ and the mutation robustness $\rho$, defined as the probability that a mutation preserves functional infectivity. In practice, mutation effects depend on epistatic interactions, genome organization, and context-dependent fitness landscapes. The framework does not attempt to model these fine-scale genetic interactions. Instead, it focuses on whether mutation can arise early enough and in sufficient quantity to influence transmission, which is the dominant constraint for long-term evolution under narrow transmission bottlenecks.

Immune interaction is also simplified. Host immunity is represented as an effective reduction in replication success $R$, rather than as a dynamic, multi-layered system involving innate signaling, adaptive memory, spatial heterogeneity, and host-to-host variability. This treatment is appropriate for comparing evolutionary strategies across viruses but cannot capture transient immune dynamics, immune imprinting, or short-term immunological feedback. Consequently, the framework should be viewed as complementary to, rather than a replacement for, detailed within-host immune models.

Environmental decay is modeled as an approximately first-order process governed by a single effective decay constant. While this assumption is well supported by experimental studies across many viruses and environments, real-world conditions often involve heterogeneous microenvironments, intermittent shielding by organic material, and spatially varying stressors. These effects can introduce deviations from simple exponential decay over extended timescales. The present framework captures the dominant trend rather than fine-scale deviations.

Cross-species transmission and zoonotic emergence are not explicitly modeled. Host switching often places viruses far from optimal regions of parameter space, producing transient maladaptation followed by rapid selection. While the same hazard, mutation, and timing constraints still apply, the framework does not describe the transient dynamics of adaptation following a host jump. Extending the model to incorporate host-specific hazard surfaces would be required to address emergence in detail.

Finally, the framework is not predictive in a narrow numerical sense. It does not forecast case counts, outbreak sizes, or evolutionary trajectories. Its predictions are structural rather than quantitative: which strategies are viable, which are unstable, and which are effectively inaccessible given hazard rates, mutation robustness, and transmission bottlenecks. This limitation is deliberate. The framework aims to bound what is possible rather than to predict precisely what will occur.

Despite these limitations, the abstraction is also the framework’s principal strength. By focusing on hazard accumulation, replication timing, mutation feasibility, and transmission bottlenecks, it exposes invariant trade-offs that apply across virus families, environments, and hosts. Future extensions could integrate this constraint-based approach with molecular stability models, immune simulations, and epidemiological dynamics. In particular, combining hazard-based feasibility analysis with genomic surveillance may enable early identification of approaching mutational collapse, constrained evolutionary regimes, or transitions between transmission strategies.

\section{Conclusion: Viral Evolution Under Hazard and Timing Constraints}

Viruses display extraordinary diversity in structure, transmission mode, immune interaction, and evolutionary behavior. Yet this diversity does not arise from unconstrained adaptation. Instead, it reflects optimization under a small set of dominant physical and biological constraints: irreversible environmental decay, replication timing within hosts, mutation generation capacity, and transmission bottlenecks. By treating virions as multi-component physical systems subject to environmental failure and viruses as replicators constrained by immune pressure and mutation robustness, this work reframes viral evolution as a problem of constrained feasibility rather than descriptive classification.

The central result of the framework is that transmission fitness is governed by the balance between effective replication success and environmental decay. Replication within hosts and survival outside hosts are not independent axes of success but opposing pressures linked through structural complexity and failure pathways. Increasing immune evasion, regulatory control, or persistence within hosts necessarily introduces additional fragile components and raises the aggregate hazard rate of environmental failure. Conversely, minimizing environmental decay constrains immune interaction, mutation timing, and regulatory sophistication. Viral strategies therefore occupy narrow, structurally determined regions of parameter space rather than arbitrary points in trait space.

Within this hazard-based view, familiar patterns in virology emerge as necessary outcomes rather than historical contingencies. Environmentally transmitted and airborne viruses are structurally simple, chemically robust, and rely on replication volume rather than immune suppression or early selection. Structurally complex viruses sacrifice environmental persistence to encode immune-modulatory machinery, latency, or persistent replication. Highly adaptive viruses operate near critical boundaries where mutation can occur early enough to enable immune escape without triggering widespread loss of function. Seasonality, latency, vaccine durability, and long-term coexistence follow directly from these constraints on decay, mutation timing, and selection.

The framework also clarifies the limits of intervention. Some viruses are eradicable because narrow transmission bottlenecks, low mutation robustness, and early exponential replication make adaptive escape statistically inaccessible. Others are intrinsically endemic because mutation is generated early and selection acts continuously within and across hosts, ensuring persistent diversification despite immune pressure. Environmental interventions succeed only when environmental decay constitutes the dominant bottleneck, while vaccination succeeds only when immune pressure suppresses replication before adaptive variants can arise and be transmitted. These outcomes are not matters of policy preference but consequences of physical decay, biological fragility, and evolutionary timing.

By unifying environmental hazard, replication dynamics, mutation feasibility, and transmission architecture within a single constraint-based model, this work provides a principled foundation for understanding viral diversity and long-term behavior. It does not replace molecular, immunological, or epidemiological models; rather, it constrains them, identifying which evolutionary pathways are viable and which are fundamentally inaccessible. Ultimately, viral evolution is neither arbitrary nor infinitely plastic. It is shaped by failure rates, bottlenecks, and timing. Recognizing these constraints transforms viral diversity from a catalog of exceptions into a coherent landscape governed by first principles, offering clearer insight into both the behavior of existing viruses and the limits of those yet to emerge.

\bibliographystyle{unsrt}
\bibliography{references}

\end{document}